\documentclass[onecolumn]{aastex631}

\shorttitle{magnetic field in G28.34}
\shortauthors{Hwang et al.}

\begin{document}

\title{The JCMT BISTRO-3 Survey: Variation of magnetic field orientations on parsec and sub-parsec scales in the massive star-forming region G28.34$+$0.06}

\author[0000-0001-7866-2686]{Jihye Hwang}
\email{astrojhwang@gmail.com}
\affil{Korea Astronomy and Space Science Institute (KASI), 776 Daedeokdae-ro, Yuseong-gu, Daejeon 34055, Republic of Korea}
\affil{Institute for Advanced Study, Kyushu University, Japan}
\affil{Department of Earth and Planetary Sciences, Faculty of Science, Kyushu University, Nishi-ku, Fukuoka 819-0395, Japan}

\author[0000-0002-8557-3582]{Kate Pattle}
\affil{Department of Physics and Astronomy, University College London, WC1E 6BT London, UK}

\author[0000-0002-3179-6334]{Chang Won Lee}
\affil{Korea Astronomy and Space Science Institute (KASI), 776 Daedeokdae-ro, Yuseong-gu, Daejeon 34055, Republic of Korea}
\affil{University of Science and Technology, Korea (UST), 217 Gajeong-ro, Yuseong-gu, Daejeon 34113, Republic of Korea}

\author[0000-0001-5996-3600]{Janik Karoly}
\affil{Department of Physics and Astronomy, University College London, WC1E 6BT London, UK}

\author[0000-0003-2412-7092]{Kee-Tae Kim}
\affil{Korea Astronomy and Space Science Institute (KASI), 776 Daedeokdae-ro, Yuseong-gu, Daejeon 34055, Republic of Korea}
\affil{University of Science and Technology, Korea (UST), 217 Gajeong-ro, Yuseong-gu, Daejeon 34113, Republic of Korea}

\author[0000-0002-1229-0426]{Jongsoo Kim}
\affil{Korea Astronomy and Space Science Institute (KASI), 776 Daedeokdae-ro, Yuseong-gu, Daejeon 34055, Republic of Korea}

\author[0000-0002-4774-2998]{Junhao Liu}
\affil{Division of ALMA, National Astronomical Observatory of Japan, Mitaka, Tokyo 181-8588, Japan}

\author[0000-0002-5093-5088]{Keping Qiu}
\affil{School of Astronomy and Space Science, Nanjing University, 163 Xianlin Avenue, Nanjing 210023, People's Republic of China}
\affil{Key Laboratory of Modern Astronomy and Astrophysics (Nanjing University), Ministry of Education, Nanjing 210023, People's Republic of China}

\author[0000-0002-9907-8427]{A-Ran Lyo}
\affil{Korea Astronomy and Space Science Institute (KASI), 776 Daedeokdae-ro, Yuseong-gu, Daejeon 34055, Republic of Korea}

\author[0000-0002-5881-3229]{David Eden}
\affil{Department of Physics, University of Bath, Claverton Down, Bath BA2 7AY, UK}
\affil{Armagh Observatory and Planetarium, College Hill, Armagh BT61 9DB, UK}

\author[0000-0003-2777-5861]{Patrick M. Koch}
\affil{Academia Sinica Institute of Astronomy and Astrophysics, No.1, Sec. 4., Roosevelt Road, Taipei 10617, Taiwan}

\author[0000-0002-1959-7201]{Doris Arzoumanian}
\affil{Institute for Advanced Study, Kyushu University, Japan}
\affil{Department of Earth and Planetary Sciences, Faculty of Science, Kyushu University, Nishi-ku, Fukuoka 819-0395, Japan}

\author[0000-0002-4541-0607]{Ekta Sharma}
\affil{CAS Key Laboratory of FAST, National Astronomical Observatories, Chinese Academy of Sciences, People's Republic of China}

\author[0000-0002-5391-5568]{Fr\'ed\'erick Poidevin}
\affil{Instituto de Astrofis\'{i}ca de Canarias, E-38205 La Laguna, Tenerife, Canary Islands, Spain}
\affil{Departamento de Astrof\'{\i}sica, Universidad de La Laguna (ULL), E-38206 La Laguna, Tenerife, Spain}

\author[0000-0002-6773-459X]{Doug Johnstone}
\affil{NRC Herzberg Astronomy and Astrophysics, 5071 West Saanich Road, Victoria, BC V9E 2E7, Canada}
\affil{Department of Physics and Astronomy, University of Victoria, Victoria, BC V8W 2Y2, Canada}

\author[0000-0002-0859-0805]{Simon Coud\'{e}}
\affil{Department of Earth, Environment, and Physics, Worcester State University, Worcester, MA01602, USA}
\affil{Center for Astrophysics | Harvard \& Smithsonian, 60 Garden Street, Cambridge, MA02138, USA}

\author[0000-0001-8749-1436]{Mehrnoosh Tahani}
\affiliation{Banting and KIPAC Fellow: Kavli Institute for Particle Astrophysics \& Cosmology (KIPAC), Stanford University, Stanford, CA 94305, USA}

\author[0000-0003-1140-2761]{Derek Ward-Thompson}
\affil{Jeremiah Horrocks Institute, University of Central Lancashire, Preston PR1 2HE, UK}

\author[0000-0002-6386-2906]{Archana Soam}
\affil{Indian Institute of Astrophysics, II Block, Koramangala, Bengaluru 560034, India}

\author[0000-0001-7379-6263]{Ji-hyun Kang}
\affil{Korea Astronomy and Space Science Institute (KASI), 776 Daedeokdae-ro, Yuseong-gu, Daejeon 34055, Republic of Korea}

\author[0000-0003-2017-0982]{Thiem Hoang}
\affil{Korea Astronomy and Space Science Institute (KASI), 776 Daedeokdae-ro, Yuseong-gu, Daejeon 34055, Republic of Korea}
\affil{University of Science and Technology, Korea (UST), 217 Gajeong-ro, Yuseong-gu, Daejeon 34113, Republic of Korea}

\author[0000-0003-4022-4132]{Woojin Kwon}
\affil{Department of Earth Science Education, Seoul National University, 1 Gwanak-ro, Gwanak-gu, Seoul 08826, Republic of Korea}
\affil{SNU Astronomy Research Center, Seoul National University, 1 Gwanak-ro, Gwanak-gu, Seoul 08826, Republic of Korea}
\affil{The Center for Educational Research, Seoul National University, 1 Gwanak-ro, Gwanak-gu, Seoul 08826, Republic of Korea}

\author[0000-0002-5913-5554]{Nguyen Bich Ngoc}
\affil{Vietnam National Space Center, Vietnam Academy of Science and Technology, 18 Hoang Quoc Viet, Hanoi, Vietnam}
\affil{Graduate University of Science and Technology, Vietnam Academy of Science and Technology, 18 Hoang Quoc Viet, Cau Giay, Hanoi, Vietnam}

\author[0000-0003-0014-1527]{Eun Jung Chung}
\affil{Korea Astronomy and Space Science Institute (KASI), 776 Daedeokdae-ro, Yuseong-gu, Daejeon 34055, Republic of Korea}

\author[0000-0001-7491-0048]{Tyler L. Bourke}
\affil{SKA Observatory, Jodrell Bank, Lower Withington, Macclesfield SK11 9FT, UK}
\affil{Jodrell Bank Centre for Astrophysics, School of Physics and Astronomy, University of Manchester, Oxford Road, Manchester, M13 9PL, UK}

\author[0000-0002-8234-6747]{Takashi Onaka}
\affil{Department of Astronomy, Graduate School of Science, The University of Tokyo, 7-3-1 Hongo, Bunkyo-ku, Tokyo 113-0033, Japan}

\author[0000-0002-3036-0184]{Florian Kirchschlager}
\affil{Sterrenkundig Observatorium, Ghent University, Krijgslaan 281-S9, 9000 Gent, BE}

\author[0000-0002-6510-0681]{Motohide Tamura}
\affil{Department of Astronomy, Graduate School of Science, The University of Tokyo, 7-3-1 Hongo, Bunkyo-ku, Tokyo 113-0033, Japan}
\affil{Astrobiology Center, 2-21-1 Osawa, Mitaka-shi, Tokyo 181-8588, Japan}
\affil{National Astronomical Observatory, 2-21-1 Osawa, Mitaka-shi, Tokyo 181-8588, Japan}

\author[0000-0003-2815-7774]{Jungmi Kwon}
\affil{Department of Astronomy, Graduate School of Science, The University of Tokyo, 7-3-1 Hongo, Bunkyo-ku, Tokyo 113-0033, Japan}

\author[0000-0002-4154-4309]{Xindi Tang}
\affil{Xinjiang Astronomical Observatory, Chinese Academy of Sciences, 830011 Urumqi, People's Republic of China}

\author[0000-0003-4761-6139]{Eswaraiah Chakali}
\affil{Department of Physical Sciences, Indian Institute of Science Education and Research (IISER) Mohali, Knowledge City, Sector 81, SAS Nagar 140306, Punjab, India}
\affil{Department of Physics, Indian Institute of Science Education and Research Tirupati, Yerpedu, Tirupati - 517619, Andhra Pradesh, India}

\author[0000-0002-5286-2564]{Tie Liu}
\affil{Shanghai Astronomical Observatory, Chinese Academy of Sciences, 80 Nandan Road, Shanghai 200030, China}

\author[0000-0002-0794-3859]{Pierre Bastien}
\affil{Centre de recherche en astrophysique du Qu\'{e}bec \& d\'{e}partement de physique, Universit\'{e} de Montr\'{e}al, C.P. 6128 Succ. Centre-ville, Montr\'{e}al, QC, H3C 3J7, Canada}

\author[0000-0003-0646-8782]{Ray S. Furuya}
\affil{Institute of Liberal Arts and Sciences Tokushima University, Minami Jousanajima-machi 1-1, Tokushima 770-8502, Japan}
\affil{Visiting Associate Professor at National Astronomical Observatory of Japan, 2-21-1 Osawa, Mitaka, Tokyo 181-8588, Japan}

\author[0000-0001-5522-486X]{Shih-Ping Lai}
\affil{Institute of Astronomy and Department of Physics, National Tsing Hua University, Hsinchu 30013, Taiwan}
\affil{Academia Sinica Institute of Astronomy and Astrophysics, No.1, Sec. 4., Roosevelt Road, Taipei 10617, Taiwan}

\author[0000-0002-6868-4483]{Sheng-Jun Lin}
\affil{Academia Sinica Institute of Astronomy and Astrophysics, No.1, Sec. 4., Roosevelt Road, Taipei 10617, Taiwan}

\author[0000-0002-6668-974X]{Jia-Wei Wang}
\affil{East Asian Observatory, 660 N. A'oh\={o}k\={u} Place, University Park, Hilo, HI 96720, USA}

\author[0000-0001-6524-2447]{David Berry}
\affil{East Asian Observatory, 660 N. A'oh\={o}k\={u} Place, University Park, Hilo, HI 96720, USA}

\begin{abstract}

Magnetic fields play a significant role in star-forming processes on core to clump scales. We investigate magnetic field orientations and strengths in the massive star-forming clump P2 within the filamentary infrared dark cloud G28.34+0.06 using dust polarization observations made using SCUBA-2/POL-2 on the James Clerk Maxwell Telescope as part of the B-field In STar-forming Region Observations (BISTRO) survey. We compare the magnetic field orientations at the clump scale of $\sim$2 parsecs from these JCMT observations with those at the core scale of $\sim$0.2 parsecs from archival ALMA data, finding that the magnetic field orientations on these two different scales are perpendicular to one another. We estimate the distribution of magnetic field strengths, which range from 50 to 430 $\mu$G over the clump. The region forming the core shows the highest magnetic field strength. We also obtain the distribution of mass-to-flux ratios across the clump. In the region surrounding the core, the mass-to-flux ratio is larger than 1, which indicates the magnetic field strength is insufficient to support the region against gravitational collapse. Therefore, the change in the magnetic field orientation from clump to core scales may be the result of gravitational collapse, with the field being pulled inward along with the flow of material under gravity. 
\end{abstract}

\keywords{Star Formation (1569) --- Interstellar Medium (847) --- Magnetic fields (994)}

\received{2024 September 24}
\revised{2025 April 15}
\accepted{2025 April 16}


\section{Introduction}\label{sec:intro}

Magnetic fields play an important role in star formation and the evolution of the molecular cloud by either resisting gravitational collapse or channeling flows of gas and dust (e.g., \citealt{Hwang2021, Hwang2022}, \citealt{Pattle2023}, and references therein). 
Magnetic field lines observed in the plane of the sky show ordered structures on cloud scales ($\sim$10 pc), with fields tending to be perpendicular to high-density structures and parallel to low-density structures (e.g., \citealt{Planck2016}). However, magnetic field orientations between parsec (pc) scale clumps and sub-pc scale cores are not as consistent as the large-scale fields seen in Planck observations, instead showing a bimodal distribution (e.g., \citealt{Zhang2014}). It is known that the relative importance of magnetic fields, gravity, and turbulence can affect magnetic orientations and core fragmentation in clumps (e.g., \citealt{Liu2018, Tang2019}). It is necessary to study magnetic fields on scales from clumps ($\sim$1 pc) to cores ($\sim$0.1 pc) in order to discern the role of magnetic fields in the overall star formation process (e.g., \citealt{Koch2022}). 

Magnetic fields in star-forming regions can be studied using polarized emission from dust grains aligned with respect to magnetic field lines by Radiative Alignment Torques \citep{Lazarian2007}. Polarimeters on the James Clerk Maxwell Telescope (JCMT), Stratospheric Observatory for Infrared Astronomy (SOFIA), Submillimeter Array (SMA) and Atacama Large Millimeter/submillimeter Array (ALMA) have been widely used to study magnetic field properties in star-forming regions from clouds to cores (e.g., references in \citealt{Pattle2023}). Several studies have shown the multi-scale magnetic field structures in star-forming regions (e.g., \citealt{Zhang2014, Hull2017, Koch2018, Koch2022, Liu2024}). For example, magnetic field lines in a Class 0 protostellar source in Serpens have different directions on cloud and core scales \citep{Hull2017}. Conversely, a consistent magnetic field orientation from core to clump scales is found in several star-forming regions (e.g, \citealt{Tang2019, Liu2024}), which may imply that the magnetic field dominates over gravity and turbulence. To understand the role of the magnetic field in forming cores within a cloud or clump, we therefore need to examine multi-scale magnetic field properties.

Magnetic field strengths in molecular clouds are often estimated using the Davis-Chandrasekhar-Fermi (DCF) method \citep{Davis1951, ChanFer1953}. An assumption in the method is that the underlying magnetic field is uniform, and variations are the result of distortion of the field by non-thermal gas motions. This distortion of magnetic field lines is estimated using polarization angle dispersion. However, magnetic fields may have underlying ordered variation as a result of gravitational infall, rotation, or other gas flows. In order to account for this curved magnetic field structure and estimate the angle dispersion due to non-thermal gas motions, \citet{Hwang2021} suggested a new method, extending the approach developed by \citet{Pattle2017}. They calculated polarization angle dispersion in a box smaller than the radius of curvature of the magnetic field, in which the field lines can be assumed to be uniform. Using this moving box method, a map of the magnetic field strength can be obtained. The distribution of magnetic field strengths obtained using this method can help to better constrain the magnetic field strength inferred from dust polarization observations and thus to investigate the effect of magnetic fields on core formation within a cloud or clump.

G28.34$+$0.06 (hereafter G28) is an Infrared Dark Cloud (IRDC) located at 4.8 kpc \citep{Carey1998}. IRDCs are thought to be in an early stage of the massive star formation process \citep{Wang2008}. There are two clumps, P1 and P2, in the southern and northern parts of G28, respectively (as shown in Figure \ref{fig:polmag}) which have previously been identified in NH$_3$ intensity maps \citep{Wang2008}. Recently, \citet{Liu2024} presented magnetic field orientations in G28 using JCMT and ALMA observations, showing that several cores have major axes aligned along the major axis of the P1 clump, while magnetic field lines are perpendicular to that axis. The magnetic field might regulate this aligned core fragmentation. 

However, the P2 clump appears to show a different mode of core fragmentation than that in the P1 clump \citep{Liu2020}. The core embedded in the P2 clump contains complex 15 dense condensations. \citet{Tang2019} suggested that core fragmentation types can be affected by the energy balance between gravity, turbulence, and the magnetic field. To investigate why the P2 and P1 clumps show different core fragmentation types, it is necessary to determine their energy balance. However, there are a few cases in which the role of the magnetic field in core fragmentation can be studied (e.g., \citealt{Tang2019}). G28 is a good target to examine this role due to the two distinctive core fragmentation types in clumps P1 and P2. The P1 clump has already been studied, so we focus on the P2 clump in this work.

In this paper we present the results of polarized dust continuum and N$_2$H$^+$ observations of the P2 clump in G28 made using the JCMT as part of the B-field In STar-forming Region Observations (BISTRO) survey \citep{WardThompson2017} and the 14-m telescope at the Taeduk Radio Astronomy Observatory (TRAO) as a PI project, respectively. The N$_2$H$^+$ traces dense gas with a critical density of $\sim$ 10$^5$ \citep{Shirley2015}, which is comparable to the density obtained by dust continuum. We used the velocity dispersion of N$_2$H$^+$ to estimate the magnetic field strength using the DCF method. We obtain magnetic field orientations and strengths in the P2 clump of G28 on $\sim$2 pc scales (at a distance of 4.8 kpc; \citealt{Carey1998}), and compare these with those on $\sim$0.2 pc scales measured using ALMA in the P2 clump region, and examine the relative importance of magnetic fields, gravity, and turbulence in the P2 clump on multi-spatial scales. 

This paper is organized as follows: in Section \ref{sec:obs}, we describe the polarization and molecular line observations made using the JCMT and TRAO, and the archival ALMA data for G28. We present magnetic field orientations, strengths, and properties in the P2 clump in G28 in Section \ref{sec:Res}. Measurements of mass-to-flux ratio and energy balance in the P2 clump are described in Section \ref{sec:Con}. We summarize our results in Section \ref{sec:Sum}.

\section{Observations and Data reduction} \label{sec:obs}

\subsection{JCMT Observations}

We observed G28 in linearly polarized light, arising from dust continuum emission, using the POL-2 polarimeter mounted at the SCUBA-2 bolometer camera \citep{Holland2013} on the JCMT at 450 and 850 $\micron$ simultaneously. In this work, we use the 850 $\micron$ data only. The observations were conducted as part of the BISTRO Survey (Project ID : M20AL018). The 27 data sets were obtained between September 2020 and August 2022. The duration of each observation was 31-minutes. The effective beam size of the JCMT at 850 $\micron$ is 14$''$.1 \citep{Dempsey2013}. The observations were carried out in the JCMT Weather Band 2, in which the atmospheric opacity at 225 GHz ($\tau_{225 \mathrm{GHz}}$) is between 0.05 and 0.08. The observing mode was the POL-2 DAISY scan pattern, which produces a circular map with a diameter of 11$'$. The root mean square (rms) noise values are lower and uniform in the 3$'$-diameter central region of the map and increase towards the edge of the map. To improve the rms noise value, we further used 18 datasets obtained using the JCMT between February 2022 and June 2022 under the project code M22AP018 (PI: Junhao Liu, see \citealt{Liu2024}). The data reduction procedures are explained in Appendix \ref{sec:datareduce}.


The Stokes $I$, $Q$, and $U$ maps obtained are in units of pW with a 4$''$ pixel grid. The flux conversion factor at 850 $\micron$ is 495 $\pm$ 32 Jy beam$^{-1}$ pW$^{-1}$ which is appropriate for data observed after June 30, 2018 \citep{Mairs2021}. The POL-2 data are further corrected by an additional factor of 1.35 (which results in a final conversion factor of 668 Jy beam$^{-1}$ pW$^{-1}$), due to the reduced optical throughput caused by the introduction of the spinning POL-2 half-wave plate into the SCUBA-2 light path \citep{Mairs2021}. This flux conversion factor is stable with time since 2018; although some scatter around the mean value may be seen in individual observations, our coadded set of 27 maps will be well-calibrated. The rms values of the Stokes $I$, $Q$, and $U$ maps are 7.6, 4.4, and 4.3 mJy beam$^{-1}$, respectively. We binned the polarization segments to a 12$''$ pixel grid using the $binsize$ parameter in the Starlink software; this pixel size is chosen for its similarity to the primary beam size of the JCMT at 850 $\micron$. The polarization angle ($\theta_{obs}$), debiased polarization intensities ($PI$), debiased polarization fraction ($p$), and their uncertainties are obtained using the Stokes parameters,

\begin{eqnarray}
\theta_{obs} =1/2 \arctan(U/Q),\label{eq:pol}\\
\Delta{\theta_{obs}}=0.5\sqrt{(U\Delta Q)^2+(Q\Delta U)^2}/(Q^2+U^2), \label{eq:angerr}\\
PI=\sqrt{Q^2+U^2-\Delta{PI}^2},\\
\Delta{PI}=\sqrt{((Q\Delta Q)^2+(U\Delta 
 U)^2}/(Q^2+U^2) \\
p = (Q^2 + U^2 - 0.5[(\Delta Q)^2+(\Delta U)^2])^{1/2}/I,\\
\Delta p=\sqrt{\Delta{PI}^2/I^2 + (\Delta I^2(Q^2+U^2))/I^4},
\end{eqnarray}
where $\Delta Q^2$, $\Delta U^2$, $\Delta I^2$ and   are the variances of Stokes $Q$, $U$, and $I$, respectively.

\begin{figure*}[thb!]
\epsscale{1.1}
\plotone{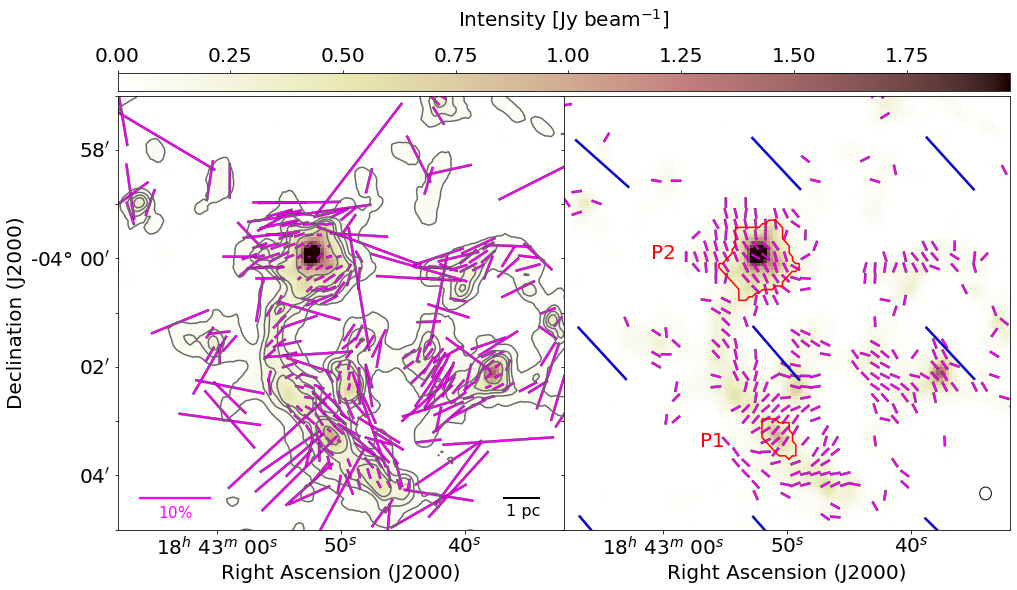}
\caption{(Left) Map of polarization segments in G28 derived from our by the JCMT POL-2 850 $\micron$ observations. The background image shows Stokes $I$ intensity. The gray contours mark intensities of 10, 30, 50, 70, and 150 $\sigma_I$, where $\sigma_I$ is the Stokes $I$ rms intensity of 7.6 mJy beam$^{-1}$. The intensity scale of the image is shown in the horizontal color bar. The selection criteria of the segments are $I$/$\Delta I \geq$ 10, $p$/$\Delta p \geq$ 3 and $p$ $<$ 20\%, where $I$, $p$, $\Delta I$ and $\Delta p$ are intensity, polarization fraction, and their uncertainties, respectively. The length of the polarization segments is shown in magenta scales with $p$. A scale bar showing $p =$ 10\% is plotted in the bottom left corner. A physical scale bar marking a distance of 1 pc is shown in the bottom right corner. (Right) Map of magnetic field segments, again shown in magenta, which are the polarization segments in the left panel rotated by 90 degrees. The segments are plotted with a uniform length for clarity. The circle in the lower right corner of the right panel indicates the JCMT beam size of 14.1$''$ at 850 $\micron$, which is equivalent to 0.32 pc at the distance of G28. Blue segments show magnetic field orientations inferred from the Planck observations. Red contours in the right panel mark clumps identified using the \textit{astrodendro} dendrogram algorithm (see Section \ref{sec:polang} for details). The two main clumps are labeled as the P1 and P2, as defined by \citet{Wang2008}.
\label{fig:polmag}}
\end{figure*}

The polarization segments that we obtained, scaled by their debiased polarization fraction, are presented in the left panel of Figure \ref{fig:polmag}. We selected polarization segments using the following criteria: $I$/$\Delta I \geq$ 10, $p$/$\Delta p \geq$ 3 and $p$ $<$ 20\% (see Figure \ref{fig:polmag}). The maximum dust polarization fraction is 20\% in \textit{Planck} observations \citep{Planck2015}.  

\subsection{TRAO Observations}

We carried out observations of N$_2$H$^+$ ($J$ = 1-0; 93.17631 GHz) spectral lines in G28 using the TRAO 14-m telescope located in Daejeon, Korea\footnote{\url{https://trao.kasi.re.kr}}. The observations were conducted from January to February 2023. The data were obtained using the on-the-fly (OTF) mapping mode of the SEcond QUabbin Optical Image Array (SEQUOIA) receiver equipped with a 4$\times$4 monolithic microwave
integrated-circuit (MMIC) preamplifier on the TRAO. Two IF modules in the observations allow us to simultaneously observe two sub-bands in a frequency range of 85 -- 100 GHz or 100 -- 115 GHz. Each spectrum has 4096 channels with a resolution of 15 kHz and a bandwidth of 62.5 MHz. The beam efficiency is 0.48 at 98 GHz \citep{Jeong2019}.

The N$_2$H$^+$ ($J$ = 1-0) and HCO$^+$ ($J$ = 1-0) observations were made using two intermediate frequency (IF) modules whose center frequencies are 93.176 and 89.189 GHz, respectively. The TRAO beam sizes at these wavelengths are 54$''$ and 56$''$, respectively. Mapping observations were made over areas of 10$' \times$ 10$'$. The central position of the mapping area is Right Ascension (J2000) = 18$^{\mathrm{h}}$ 42$^{\mathrm{m}}$ 50.9$^{\mathrm{s}}$ and Declination (J2000) = --04$^{\mathrm{\circ}}$ 03$^{\mathrm{'}}$ 14$^{\mathrm{''}}$.

We performed four OTF scan observations over the same mapping area and added the scan data to get one averaged map. The integration time for each scan of size 10$' \times$ 10$'$ was about 100 minutes and thus about 6 hours 40 minutes in total were used for the full observations. The map was reconstructed with a pixel size of 22$''$ and a spectral resolution of 0.3 km s$^{-1}$. The rms of the N$_2$H$^+$ spectra varies from 0.02 to 0.03 K. Further analysis of the map data was performed using the Continuum and Line Analysis Single-dish Software (CLASS; \citealt{Pety2005}; \citealt{gildasteam2013}).

\subsection{ALMA archive data}

We used dust polarization observations of a core embedded in the P2 clump of the G28 obtained using ALMA in Band 6. 
The receiver for the observations was tuned to cover $\sim$1.27--1.29 GHz, with a total bandwidth of 5.6 GHz in the full polarization mode. 
The ALMA observations were made between April 2017 and June 2018 as part of the projects of 2016.1.00248.S (Cycle 4; PI: Qizhou Zhang) and 2017.1.00793.S (Cycle 5; PI: Qizhou Zhang). 
The data were presented by \citet{Liu2020} and reduced using the Common Astronomy Software Applications (CASA; \citealt{McMullin2007}). The detailed data reduction processes are described by \citet{Liu2020}. The synthesized beam sizes in the Stokes $I$, $Q$, and $U$ maps are 0$''$.8 $\times$ 0$''$.6 ($\sim$0.014 pc $\times$ 0.019 pc) and the maximum recoverable scale in the observation is $\sim$7$''$ ($\sim$0.16 pc). We produced polarization segments on 0.36 $''$ $\times$ 0.36 $''$ pixels, which is about half of the beam size. 
We calculated the polarization angles of the ALMA data using Equation (\ref{eq:pol}).


\section{Results}\label{sec:Res}

\subsection{Magnetic field orientations}

The polarization segments from our JCMT/Pol2 observations are shown in the left panel of Figure \ref{fig:polmag}. In the right panel, these segments are rotated by 90 degrees to show magnetic field orientations, with all segments scaled to the same length to better show the pattern of magnetic field orientations. Magnetic field lines are well-ordered within the P1 and P2 clumps in a fashion roughly perpendicular to the major axis of each elongated clump. The blue segments represent the \textit{Planck} large-scale magnetic field orientations at 350 GHz ($\sim$857 $\micron$). The beam size for the \textit{Planck} observations is 5$'$ ($\sim$7 pc at the distance of G28), which covers the overall region of G28. The large-scale magnetic field orientation is directed from North-East to South-West.
As the details of the magnetic field orientation from cloud to core scales in the P1 clump were discussed by \citet{Liu2024}, 
 in this work, we focus on the magnetic field orientation in the P2 clump.  

\begin{figure*}[thb!]
\epsscale{1.0}
\plotone{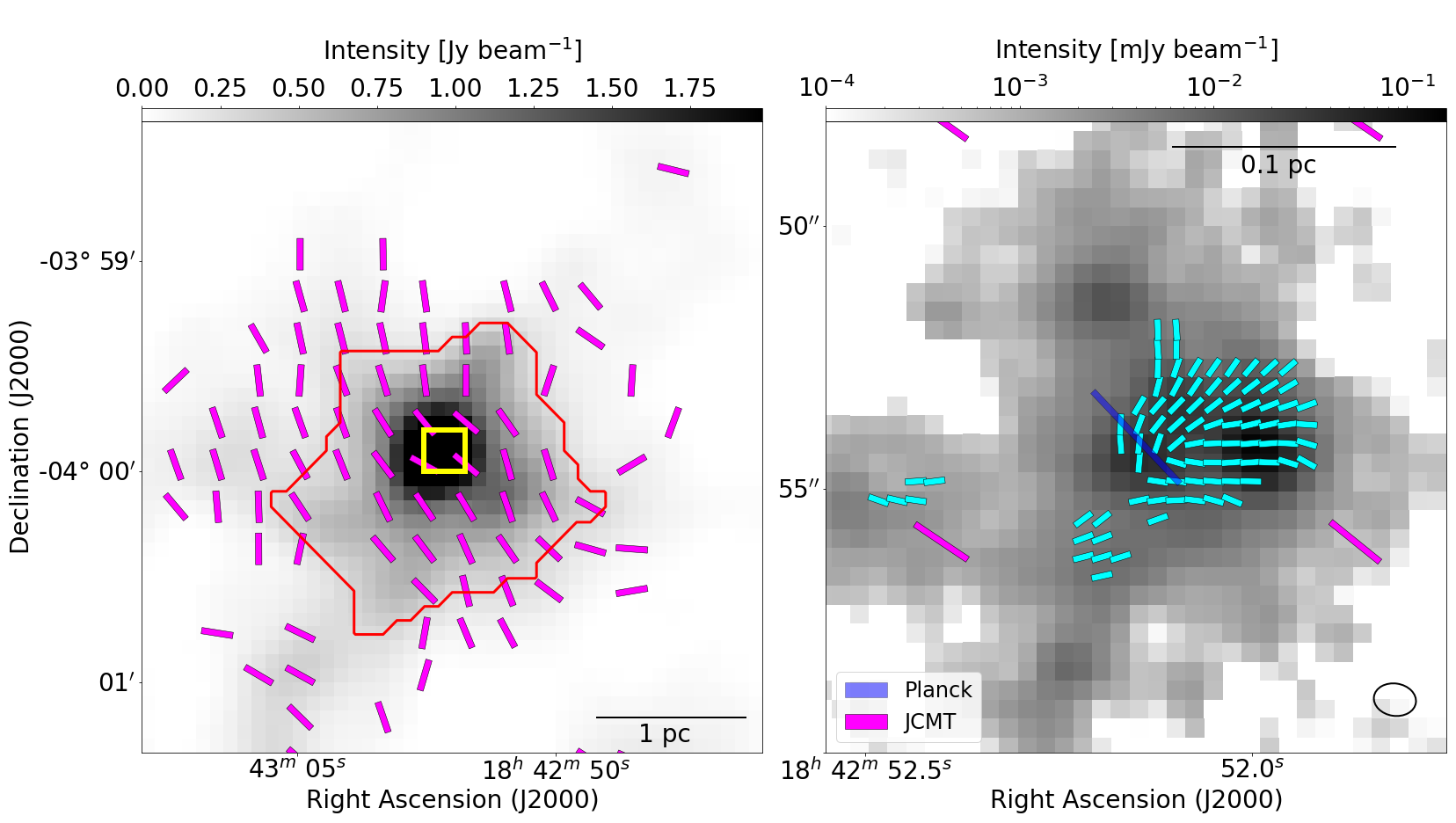}
\caption{(Left) Magnetic field orientations in the P2 clump, zoomed in from Figure \ref{fig:polmag}. The yellow rectangle marks the location of the core detected in the ALMA observations. The pink segments and red contour are the same as in Figure \ref{fig:polmag}. (Right) Magnetic field orientations in the ALMA core embedded in the P2 clump, in the region marked in yellow in the left panel. The background image shows the ALMA 1.3 mm continuum emission. The pink and blue segments show magnetic field orientations obtained using the JCMT and \textit{Planck}, respectively. Cyan segments show magnetic field orientations obtained using ALMA with $PI$/$\Delta PI$ $>$ 3. The ellipse in the lower right corner is the ALMA synthesized beam size, 0$''$.8 $\times$ 0$''$.6 (0.014 $\times$ 0.019 pc at a distance of 4.8 kpc). \label{fig:polmag2}}
\end{figure*}

\begin{figure*}[thb!]
\epsscale{0.7}
\plotone{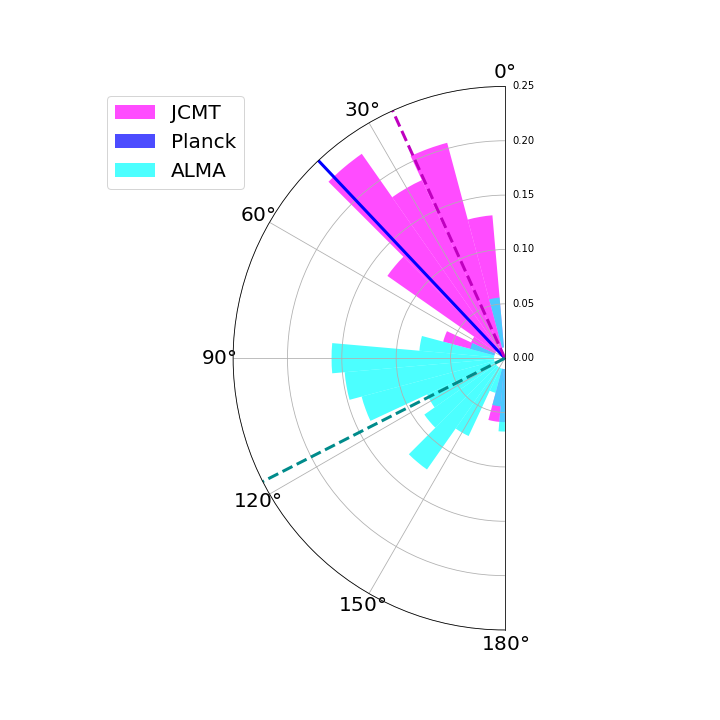}
\caption{Magnetic field orientations obtained using JCMT and ALMA are shown as magenta and cyan histograms on a polar bar chart. The \textit{Planck} magnetic field direction is marked with a blue line. The dark magenta and cyan dashed lines show the mean orientations obtained using the JCMT and ALMA, 24 and 117 degrees, respectively. The lengths of the bars represent the fraction of the polarization segments with position angles in the range of the bar. We show all segments within the area of the P2 clump bounded by the red contour shown in Figure \ref{fig:polmag}. Due to the 180-degree ambiguity on the direction of polarization segments, diametrically opposite angles on the plot agree. \label{fig:polmag3}}
\end{figure*}


The left panel of Figure \ref{fig:polmag2} shows a zoomed-in view of the magnetic field distribution in the P2 clump. The right panel of Figure \ref{fig:polmag2} shows the magnetic field distribution in the core embedded within the P2 clump; the area shown is marked with a yellow rectangle in the left panel. In both panels, pink segments show magnetic field distribution obtained using the JCMT, while cyan segments in the right panel indicate the magnetic field distribution obtained using ALMA, with selection criteria of $\mathrm{PI}/\sigma_{\mathrm{PI}}$ $>$ 3, where $\mathrm{PI}$ and $\sigma_{\mathrm{PI}}$ are the polarized intensity and its uncertainty. The field orientations on JCMT and ALMA scales are roughly perpendicular to one another. Figure \ref{fig:polmag3} is a polar chart showing the distribution of magnetic field orientations. 
Due to the 180-degree ambiguity on the direction of polarization segments, diametrically opposite angles on the plot agree. The \textit{Planck} magnetic field orientation is shown in blue.
The mean values of the JCMT (magenta histogram) and ALMA (cyan histogram) magnetic field directions are 24 and 117 degrees, with standard deviations of 19 and 32 degrees, respectively, indicating that the two magnetic field orientations are roughly perpendicular to one another with a mean angle difference of 87 degrees.

\subsection{Polarization angle dispersion} \label{sec:polang}

\begin{figure*}[thb!]
\epsscale{0.7}
\plotone{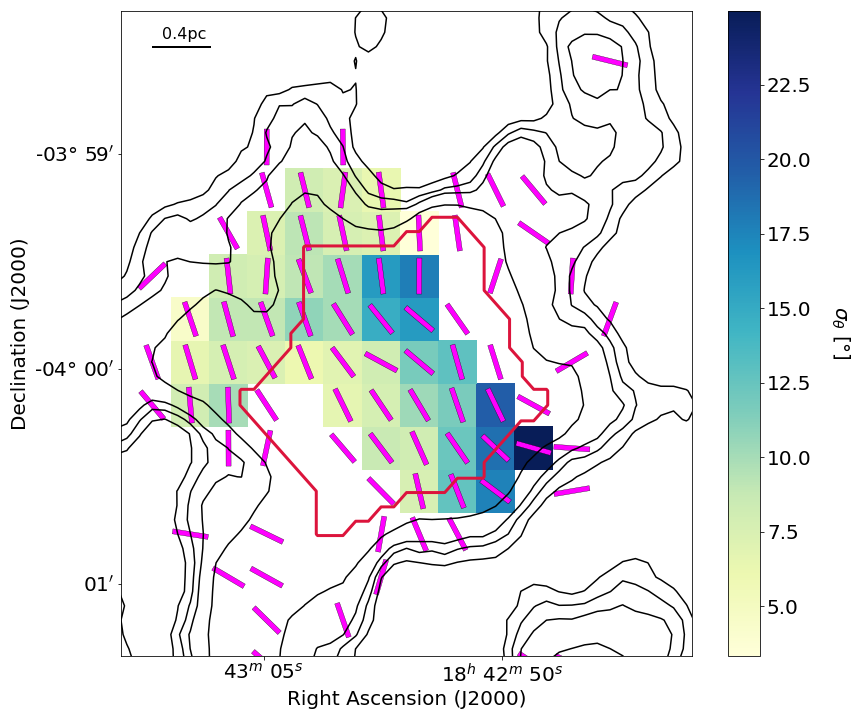}
\caption{Map of polarization angle dispersion estimated using the JCMT data in the P2 clump of G28. Pink segments are magnetic field orientations derived from the JCMT observations. The red contour is the boundary of the P2 clump, identified using the $astrodendro$ algorithm, as shown in the right panel of Figure \ref{fig:polmag}. Black contours indicate the 850 $\micron$ intensity levels of are 5, 10, 15, and 30 $\sigma_I$ from the edge to the center of the map, where $\sigma_I$ is the Stokes I rms value, 7.6 mJy beam$^{-1}$. 
\label{fig:poldisp}}
\end{figure*}

We obtained a map of polarization angle dispersion ($\sigma_\theta$) using the method introduced by \citet{Hwang2021}, in which the mean magnetic field orientations are determined within a small box region. To decide the box size, we estimated a radius of curvature for a circle on which two adjacent segments become tangents \citep{Koch2012}. We estimated the radii of curvature of four polarization segment pairs between a reference pixel and adjacent left, right, upper, and lower neighbors. We took the mean of these four radii of curvature as the radius of curvature for the polarization segment in the reference pixel. By repeating these calculations, we obtained radii of curvature for all pixels within the P2 clump. The derived radii of curvatures are larger than 36$''$, and thus we assume that the mean magnetic field orientation in a box of 36$''$ $\times$ 36$''$ (3 $\times$ 3 pixels) is uniform. In this way, we measured the mean polarization angle and angle dispersion within the box and then moved the box pixel by pixel to estimate the distribution of polarization angle dispersion (Figure \ref{fig:poldisp}) across the clump. The polarization angle dispersion in the P2 clump within the red contour marked in Figure \ref{fig:poldisp} varies from 3 to 20 degrees, with a mean value of 11 $\pm$ 4 degrees. The red contour marks the boundary of the P2 clump, which we identified using a dendrogram algorithm. For this, we used the Python package $astrodendro$\footnote{A minimum intensity level of 27 mJy beam$^{-1}$ (3 $\sigma_I$) and a minimum size of hierarchical structures of 14.1$''$ (the JCMT beam size) were used when applying the algorithm.}, which is used to recognize hierarchical structures in molecular clouds \citep{Rosolowsky2008}. The uncertainties in each pixel are estimated from the measurement errors on the polarization segments (Equation (\ref{eq:angerr})) in each box centered on the pixel.   

\subsection{Volume density}

To estimate the column density, and thus the volume density of the P2 clump we considered the area within the boundary of the P2 clump as defined in the previous section.

\begin{figure*}[thb!]
\epsscale{0.7}
\plotone{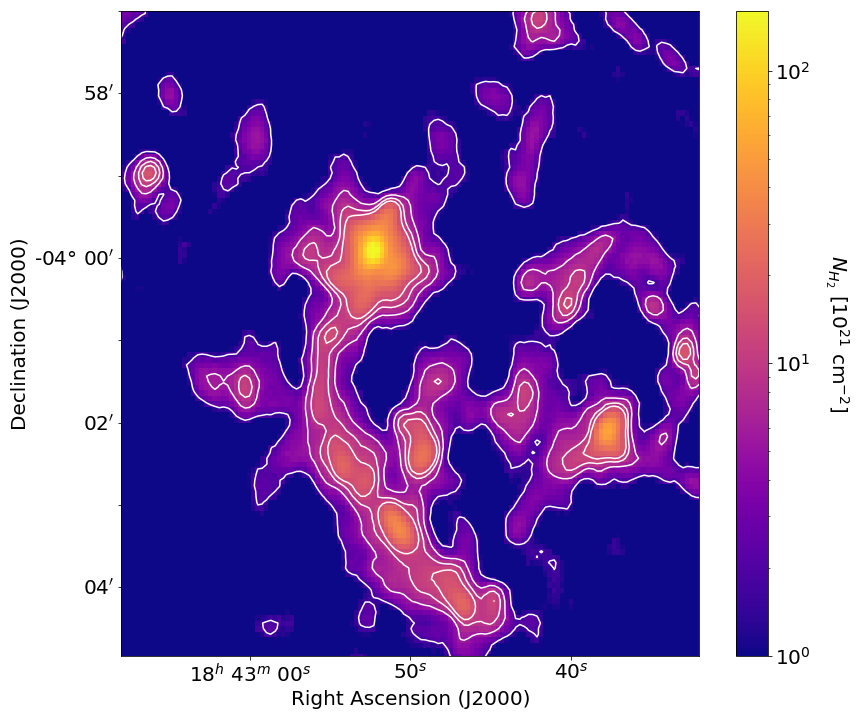}
\caption{Map of H$_2$ column density in G28. White contours indicate the 850 $\micron$ intensity levels of 10, 30, 50, and 70 $\sigma_I$. 
\label{fig:col}}
\end{figure*}

The column density of molecular hydrogen is calculated using the following equation,
\begin{equation}
N(\text{H}_2) =I_\nu/ (\mu  m_{\text{H}}  \kappa(\nu) B_\nu(T)), 
\label{eq:col}
\end{equation}
where $I_\nu$ is the intensity at a frequency $\nu$, $\mu$ = 2.8 is the mean molecular weight per particle, assuming 10\% of the total number of gas particles is helium \citep{Kauffmann2008}, $m_\mathrm{H}$ is the mass of atomic hydrogen, $\kappa(\nu)=\kappa_{\nu_0}(\nu/\nu_0)^\beta$ is the dust opacity function where $\kappa_{\nu_0}$ is 0.1 cm$^2$ g$^{-1}$ at $\nu_0 = 1$ THz, taking the dust emissivity index $\beta$ to be 2 and assuming a dust-to-gas mass ratio of 1:100 \citep{Motte2001, Andre2010}, and $B_\nu(T)$ is the Planck function at dust temperature $T$. We assumed that the dust temperature is similar to the rotation temperature of 29 K obtained using NH$_3$ ($J$, $K$) = (1,1) and (2,2) observations in the P2 clump \citep{Wang2008}. This value is comparable to dust temperatures of 19-50 K derived using the ratio of intensities at 450 and 850 $\micron$ in the clump and $\beta$ = 1.5-2.0 \citep{Carey2000}. We estimated H$_2$ column densities in each pixel within G28 using 850 $\micron$ intensity (Figure \ref{fig:col}). The mean value that we obtain is 3.0 $\times$ 10$^{22}$ cm$^{-2}$ which is comparable to the value estimated by \citet{Carey2000}, 5.7 $\times$ 10$^{22}$ cm$^{-2}$ within a factor of $\sim 2$. We also compared our results with column density maps of G28 obtained using the point process mapping (PPMAP) tool and Spectral Energy Distribution (SED) fitting by \citet{Marsh2017} and \citet{Lin2017}. Their mean column density values in the P2 clump are 11.0 $\times$ 10$^{22}$ cm$^{-2}$ and 4.7 $\times$ 10$^{22}$ cm$^{-2}$, respectively, which are consistent within a factor of a few with our value. 

To estimate the volume density from the column density of the P2 clump, we took its depth to be the equivalent radius of the P2 clump, which is the radius of a circle with an area equal to that enclosed by the red contour defining the clump (Figure \ref{fig:poldisp}). The estimated radius was $\sim$1 pc. By dividing our column densities by this radius, we obtained volume densities of molecular hydrogen in each pixel, 
\begin{equation}
n(\text{H}_2)=N(\text{H}_2)/R,    
\end{equation}
where $R$ is the equivalent radius of the P2 clump. Volume densities in the clump range between (2.9 -- 58.1) $\times$ 10$^3$ cm$^{-3}$, with a mean value of 1.0 $\times$ 10$^4$ cm$^{-3}$. 
We note that we have assumed that P2 clump is an oblate structure. If we instead took the depth to be 2$R$, the diameter of the equivalent-area circle described above, the volume density and magnetic field strength will decrease by a factor 2 and 1.4, respectively. 

\subsection{Velocity dispersion} \label{sec:veldisp}

\begin{figure*}[htb!]
\epsscale{1.1}
\plotone{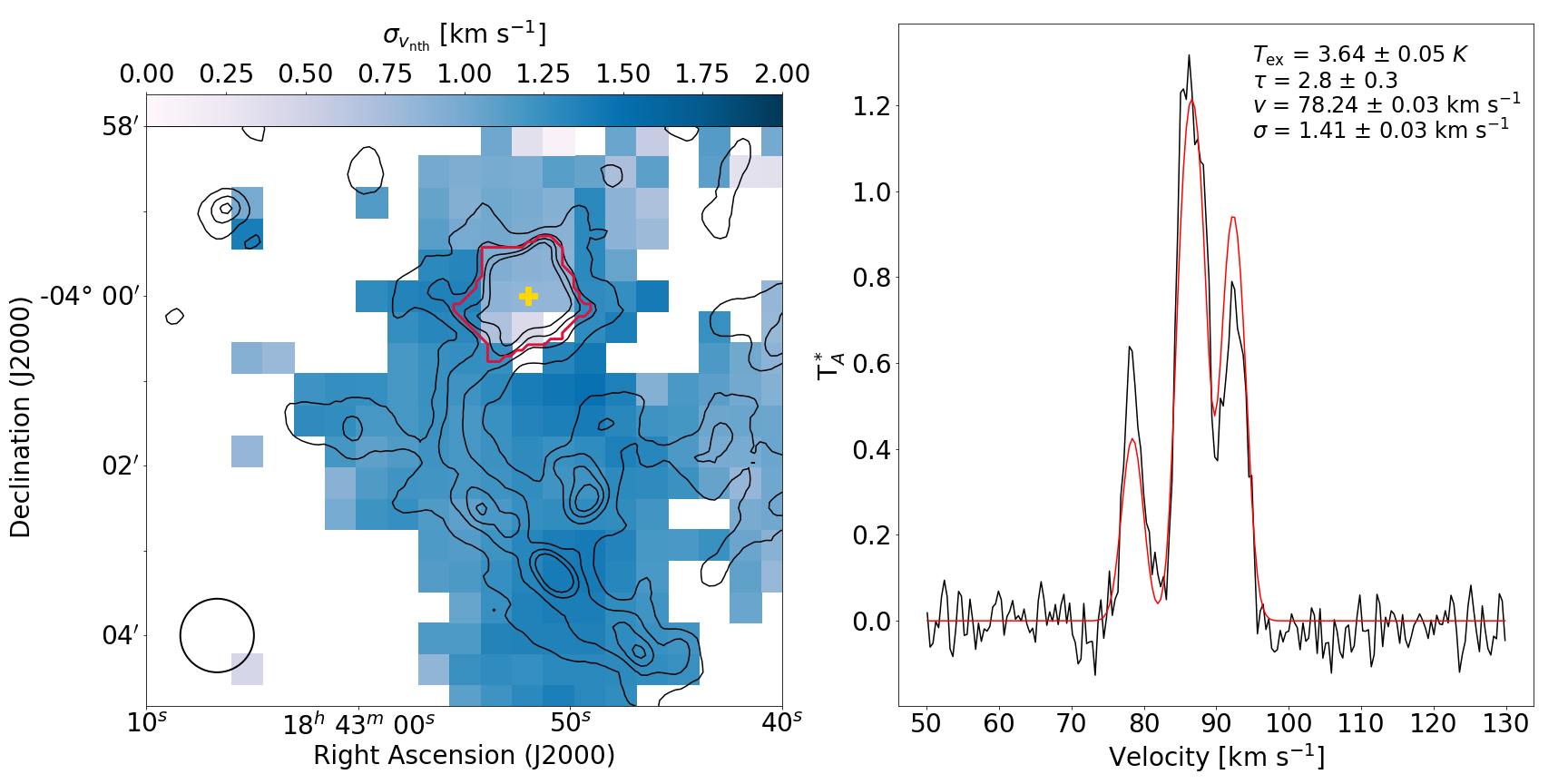}
\caption{(Left) Map of the non-thermal component of the velocity dispersion of the N$_2$H$^+$ spectral line. Black contours indicate the 850 $\micron$ intensity of 10, 30, 50, and 70 $\sigma_I$ from outside to the center. The circle in the lower left corner shows the beam size of the TRAO, 52$''$. (Right) The spectrum at the position of the yellow cross is marked in the left panel. The red line is the best-fit hyperfine structure model for the  N$_2$H$^+$ data, shown in black. The fitting results are shown in the upper right corner, where T$_{\text{ex}}$, $\tau$, $v$, and $\sigma$ are the excitation temperature, optical depth, velocity, and velocity width temperature, respectively. 
\label{fig:veldisp}}
\end{figure*}

We used N$_2$H$^+$ ($J$=1--0) spectral line data to estimate non-thermal velocity dispersion values for the DCF method. The N$_2$H$^+$ line traces dense regions, due to its high critical density of 1.5 $\times$ 10$^5$ cm$^{-3}$, and its low degree of depletion in cold dense gas (e.g., \citealt{Bergin1997, Caselli2002}). Due to these properties, it is one of the best tracers with which to measure non-thermal velocity dispersion within the P2 clump. 
The N$_2$H$^+$ ($J$ = 1--0) spectral feature have seven hyperfine structures (HFSs). However, six of these HFSs are found to be blended in G28 due of their large line widths (right panel of Figure \ref{fig:veldisp}). The observed N$_2$H$^+$ spectra show three Gaussian components. The two components in the center of the spectrum and at higher velocity each contain three hyperfine lines. Only the leftmost, low-velocity, component has one hyperfine line.  

 Figure \ref{fig:veldisp} shows a map of the non-thermal velocity dispersion obtained using the TRAO N$_2$H$^+$ observations. The beam size of the TRAO is 54$''$. However, the velocity dispersion map on a grid of 22$''$ as the spectra were observed on that grid. We fit each N$_2$H$^+$ spectrum with a hyperfine line model using the Python package of \textit{pyspeckit} (an example is shown in the right panel of Figure \ref{fig:veldisp}). We estimated the velocity dispersion from the hyperfine structure fitting. 

The non-thermal component $\sigma_{v_\mathrm{nth}}$of the velocity dispersion was estimated by subtracting the thermal component, 
\begin{equation}
\sigma_{v_\mathrm{nth}} =\sqrt{ \sigma_{v_{\text{obs}}}^2 - \frac{ k T_k}{m_{\text{N}_2\text{H}^+}}},
\label{eq:vel}
\end{equation}
where $\sigma_{v_{\text{obs}}}$ is the velocity dispersion obtained from the fit of the observed N$_2$H$^+$ spectral line, $T_k$ is the kinetic temperature which is taken here to be the rotational temperature of NH$_3$ spectral line, 29 K, and $m_{\text{N}_2\text{H}^+}$ is the mass of the N$_2$H$^+$ molecule. The velocity dispersion of the non-thermal gas component ranges from 1.29 to 1.48 km s$^{-1}$ in G28 with a mean value of 1.40 $\pm$ 0.03 km s$^{-1}$. The uncertainties are the propagation of the errors on the hyperfine structure fitting.  

Since the beam size of the TRAO is four times larger than that of the JCMT, the velocity dispersion could be overestimated. However, the non-thermal velocity dispersion determined from NH$_3$ observations made using the Very Large Array (VLA) with a beam size of 3$''$ $\times$ 5$''$ is about 1.8 km s$^{-1}$ in the P2 clump, which is comparable to our results \citep{Wang2008}. Therefore, it seems likely that the large beam of the TRAO does not affect the line broadening, so we used the non-thermal velocity dispersion obtained by the TRAO data.

\subsection{Magnetic field strengths} \label{sec:mag}

Plane-of-sky magnetic field strengths are typically inferred from dust polarization observations using the DCF method, in which we assume the magnetic field lines to be distorted by non-thermal gas motions. We used the polarization angle dispersion, the volume density and the non-thermal gas velocity dispersion to infer a map of the plane-of-sky magnetic field strength using the following simple formulation of the DCF method \citep{Crutcher2004}, 
\begin{equation} 
B_{\text{pos}} = Q\sqrt{4\pi\rho}  \frac{\sigma_{v_\mathrm{nth}}}{\sigma_\theta} \approx 9.3\sqrt{n(\text{H}_2)}\frac{\Delta V}{\sigma_\theta},
\label{eq:cf}
\end{equation}
where $B_{\text{pos}}$ is the magnetic field strength in the plane of the sky in $\mu$G, $Q$ is a correction factor which is taken to be 0.5 following \citet{Ostriker2001}, $\rho$ = $\mu m_\mathrm{H} n(\mathrm{H}_2)$ is the mass density, $\sigma_{v_\mathrm{nth}}$ is the non-thermal velocity dispersion, $\sigma_\theta$ is the polarization angle dispersion, $n(\text{H}_2)$ is the number density of molecular hydrogen in units of cm$^{-3}$, and $\Delta V = \sqrt{8 \ln 2}\sigma_v$ is the full width at half maximum (FWHM) of the non-thermal line width in units of km s$^{-1}$. We note that $Q$ = 0.5 is estimated in a cloud with $n(\mathrm{H}_2) = 100$ cm$^{-3}$ and for a length ($l$) of 8 pc. In the case of the P2 clump, the volume density and length are 10$^4$ cm$^{-3}$ and 1 pc, respectively. \citet{Liu2022} has suggested a $Q$ value of 0.28 in a cloud with $n(\mathrm{H}_2) = 10^4-10^5$ cm$^{-3}$ and $l = 0.1-1$ pc using numerical simulations. If we use this $Q$ value, the estimated magnetic field strength will decrease by a factor of $\sim$2.   

\begin{figure*}[htb!]
\epsscale{0.8}
\plotone{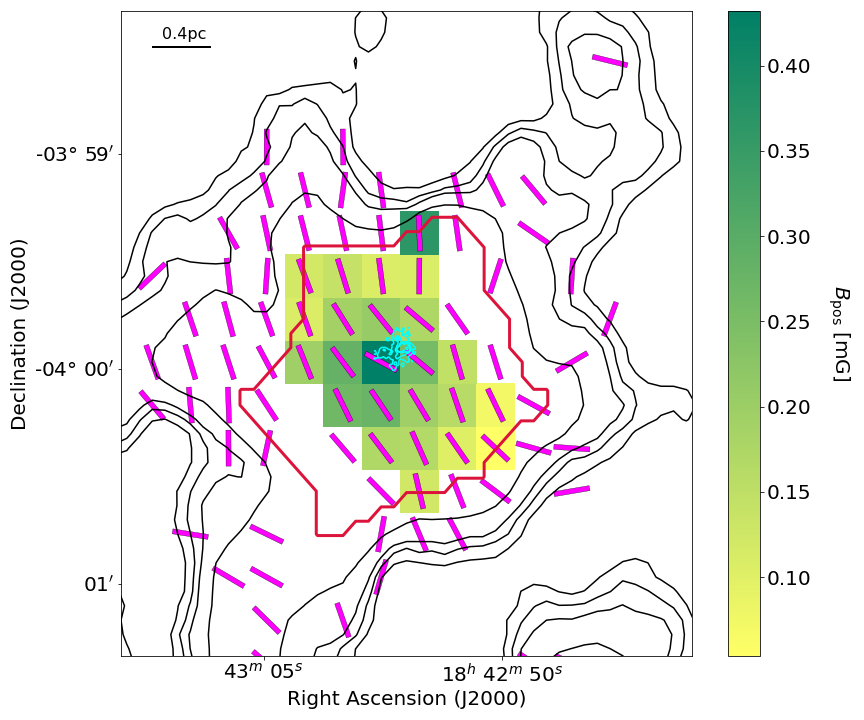}
\caption{Map of the magnetic field strength in the P2 clump of G28. Black contours are the same as in Figure \ref{fig:poldisp}. Cyan contours trace dust emission obtained using ALMA. 
\label{fig:mag}}
\end{figure*}

We obtained the map of plane-of-sky magnetic field strengths using the maps of polarization angle dispersion, volume density, and velocity dispersion. We re-grid the map to have the same coordinates as the polarization angle dispersion map with the pixel size of 12 $''$.  We then substituted these values into the equation \ref{eq:cf}. Figure \ref{fig:mag} shows the distribution of plane-of-sky magnetic field strengths in the P2 clump obtained using this method, which varies from 96 to 772 $\mu$G with a mean value of 330 $\pm$ 115 $\mu$G. If we use a $Q$ value of 0.28, the magnetic field strength will vary from 54 to 432 $\mu$G with a mean value of 185 $\pm$ 65 $\mu$G. We will use the magnetic field strengths determined using $Q$ = 0.28 in our analysis in the upcoming sections.  

We further note that the DCF method gives us the plane-of-sky component of the magnetic field strength only.  Statistically, $B_{\text{pos}}$ = ($\pi$/4)$B$ \citep{Crutcher2004}. However, this correction is only valid over an ensemble of measurements, and so we do not apply it to our value of $B_{\text{pos}}$. However, we also note that this indicates that the correction from $B_{\text{pos}}$ to $B$ is typically small ($B = 1.27 B_{\text{pos}}$), and should not alter the conclusions of our work.

We note that the DCF method is subject to significant statistical and systematic uncertainties (e.g. \citealt{Pattle2023}). Our error bars should be understood as an expression of the statistical uncertainty on the result arising from the velocity and angular dispersions, rather than as 1-sigma error bars, since there are significant systematic uncertainties on both gas density and on $Q$.  Recent reviews and meta-analyses of DCF measurements (e.g. \citealt{Liu2021, Pattle2023}) suggest that the DCF method produces magnetic field strength values that are correct to within a factor of a few. This is sufficient to inform our energetics analysis and discussion, below.

\begin{deluxetable*}{cccccc}
\tablecaption{Parameters used for estimating magnetic field strength and mass-to-flux ratio \label{tab:dcf}}
\tablecolumns{6}
\tablenum{1}
\tablewidth{0pt}
\tablehead{\colhead{$\sigma_\theta$} &
\colhead{$\sigma_{v_\mathrm{nth}}$} & \colhead{N$_{\mathrm{H}_2}$} &  \colhead{$n_{\mathrm{H}_2}$} & \colhead{$B_\mathrm{pos}$} & \colhead{$\mu_\Phi$}\\
\colhead{[degree]} & [km s$^{-1}$] & [10$^{22}$ cm$^{-2}$] &  [10$^4$ cm$^{-3}$] & [$\mu$G] & \colhead{}} 
\startdata
11 $\pm$ 4 & 1.40 $\pm$ 0.03 & 3.5 & 1.0 & 185 $\pm$ 65 & 1.5 $\pm$ 0.5 \\
\enddata
\tablecomments{These values are averaged in the P2 clump, bounded by the red contour marked on Figure \ref{fig:polmag2}. We were unable to determine quantitative uncertainties of the column density and number densities of molecular hydrogen due to the uncertainties on the flux loss, dust opacity, and depth of the P2 clump. The magnetic field strength was obtained using the DCF method with $Q$ = 0.28.}
\end{deluxetable*}

\section{Discussions}\label{sec:Con}

\subsection{Mass-to-flux ratio} \label{sec:mtob}

Mass-to-magnetic flux ratios ($\mu_\Phi$) have been used in previous studies to determine whether magnetic fields can support a star-forming region against gravitational collapse \citep{Mouschovias1976, Crutcher2004}. The observed mass-to-flux ratio is scaled with a critical value, usually taken to be the value of a magnetized disk, $1/2\pi G^{1/2}$ where $G$ is the gravitational constant \citep{Nakano1978}. The mass-to-flux ratio in units of the critical value \citep{Crutcher2004} is

\begin{equation}
\mu_\Phi = 7.6 \times10^{-21} \frac{N(\text{H}_2)}{B},\label{eq:mass_flux_ratio}      
\end{equation}

\noindent
where $B$ is the three-dimensional (3D) magnetic field strength in units of $\mu$G. A value of $\mu_\Phi$ less than one implies that a star-forming region is magnetically subcritical, in which state the magnetic field is sufficiently strong to resist gravitational collapse. A value of $\mu_\Phi$ larger than one means that a star-forming region is magnetically supercritical, in which state the magnetic field cannot prevent gravitational collapse. 

\begin{figure*}[htb!]
\epsscale{1.0}
\plotone{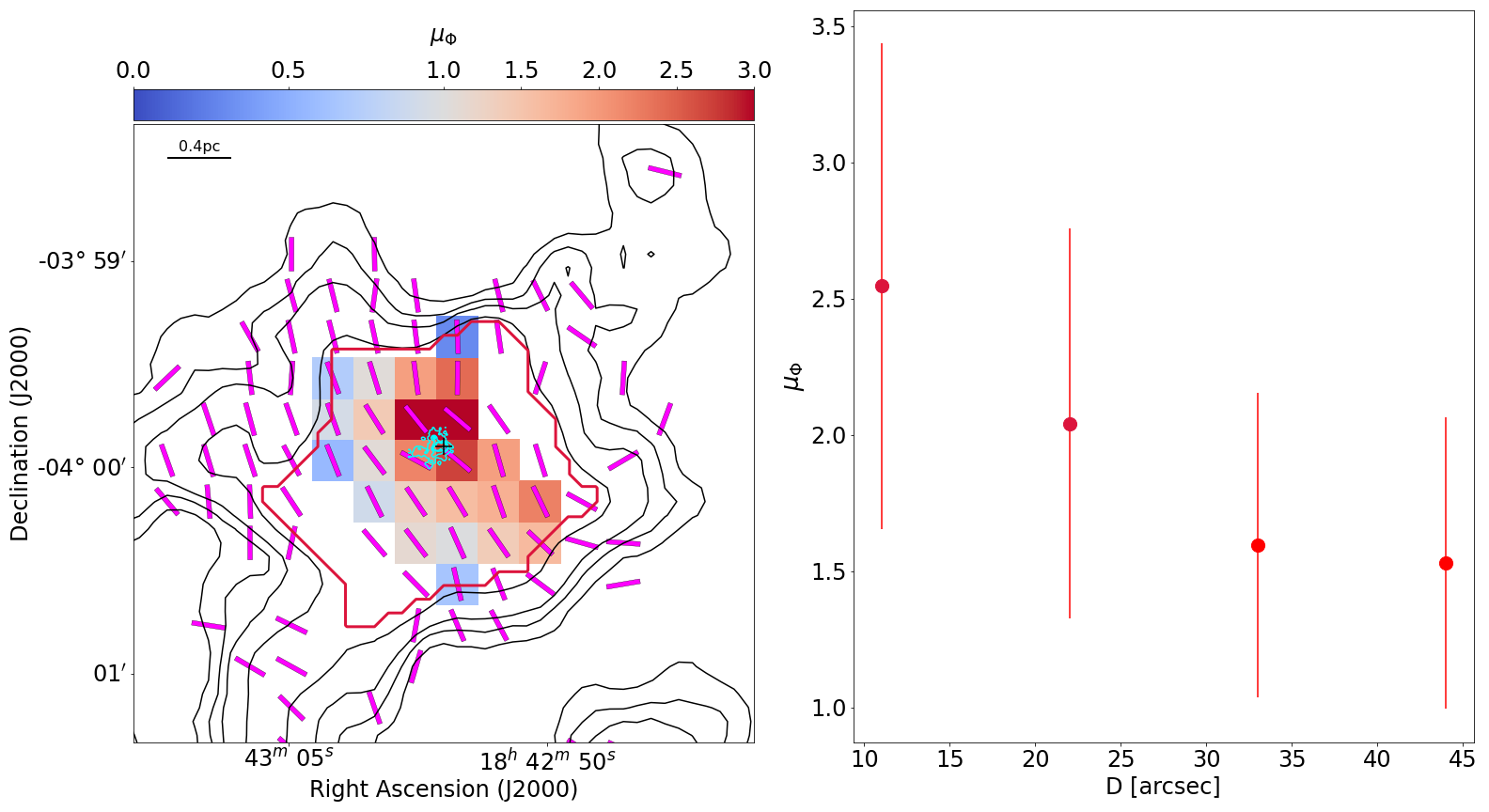}
\caption{(Left) Map of mass-to-flux ratio in the P2 clump of G28. Black and cyan contours are as described in Figure \ref{fig:mag}. The black cross marked in the center is the position of peak intensity at 850 $\micron$. (Right) The averaged mass-to-flux ratios within circular regions of given radii around the peak intensity position. 
\label{fig:mtob}}
\end{figure*}

Figure \ref{fig:mtob} shows the distribution of mass-to-flux ratio in units of the critical value, as derived from our JCMT observations. When calculating the mass-to-flux ratio, we used the magnetic field strengths in the plane of the sky obtained as described in Section \ref{sec:mag} rather than the 3D magnetic field strengths. Therefore, the estimated mass-to-flux ratios may be upper limits. The derived mass-to-flux ratios vary from 0.3 to 3.2, with a mean value of 1.5 $\pm$ 0.5.
We calculated the mean mass-to-flux ratios within circular regions of increasing radius, centered on the peak intensity position in our 850 $\micron$ Stokes $I$ map. The right panel of Figure \ref{fig:mtob} shows the estimated mean mass-to-flux ratios as a function of radius. We find that the mass-to-flux ratio decreases as the radius increases. The P2 clump is in a magnetically supercritical state over all of the radii that we consider, indicating that the magnetic field cannot support the clump against gravitational collapse. This supports the first possibility discussed in Section \ref{sec:sce}, below: that the gravitational collapse of the P2 clump may result in the change of the magnetic field orientations from clump to core scales seen when comparing JCMT and ALMA observations. 
Despite our large uncertainties, we can have some confidence that G28 is magnetically supercritical.  The DCF method is known to typically overestimate magnetic field strengths; statistically, this overestimation appears to be by a factor $\sim$3-5 \citep{Pattle2023}, although the applicability of this factor to any individual measurement is unclear. Finding a supercritical mass-to-flux ratio using the DCF method despite its tendency to overestimate magnetic field strengths thus strongly suggests that G28 is indeed supercritical. Moreover, the G28 clump is demonstrably a site of active star formation, and so our finding of a supercritical mass-to-flux ratio is fully consistent with the star formation history of the region.



\subsection{Energy Balance}

We estimate the kinetic ($E_{\rm K}$), gravitational ($E_{\rm G}$), and magnetic energies ($E_{\rm B}$) in the P2 clump using the relations,
\begin{eqnarray}
    E_{\rm K}=\frac{3}{2}M\sigma_{v_{\text{tot}}}^2, \\
    E_{\rm G}=-\frac{3}{5}\frac{GM^2}{R}, \, \\
    E_{\rm B}=\frac{1}{2}Mv_A^2,
\end{eqnarray}
where $M$ is the mass of the P2 clump, $\sigma_{v_{\text{tot}}}$ is the total velocity dispersion estimated as $\sigma_{v_{\text{tot}}}=\sqrt{\sigma_{v_\mathrm{nth}}^2+\sigma_{v_\mathrm{th}}}$ where $\sigma_{v_\mathrm{th}}$ is the thermal velocity dispersion for free particle with the mean molecular weight, $\mu$=2.37, $G$ is the gravitational constant, $R$ is the equivalent radius of the P2 clump, and $v_A$ is the Alfv\'en velocity ($B/\sqrt{4\pi\rho}$, where $\rho$ = $\mu m_\mathrm{H} n(\mathrm{H}_2)$ is the mass density).
The mass of the P2 clump is calculated using the formulation \citep{Hildebrand1983},
\begin{equation}
M = \sum_{i=1}^N \frac{I_{850,i}A}{\kappa_{\nu} B_{\nu} (T)},
\end{equation}
where $N$ is the number of pixels within the red contour defining the area of the P2 clump (Figure \ref{fig:mtob}), $I_{850,i}$ is the 850 $\micron$ flux density of $i$-th pixel, $A$ is the pixel area, and  $\kappa(\nu)$ and $B_\nu(T)$ are as defined in equation (\ref{eq:col}). We calculate a mass for the P2 clump of 2037 M$_\odot$. The mean velocity dispersion of 1.2 km s$^{-1}$ and the magnetic field strength of 185 $\mu$G in the P2 clump are used when calculating $E_{\rm K}$ and $E_{\rm B}$.

The estimated kinetic, gravitational, and magnetic energies of the P2 clump are 12 $\times$ 10$^{46}$, 21 $\times$ 10$^{46}$, and 6 $\times$ 10$^{46}$ erg, respectively. We emphasize that these values are order-of-magnitude estimates only, and are subject to significant systematic and statistical uncertainties (e.g. \citealt{Pattle2017}), but their relative values can nonetheless inform our discussion of the region. The gravitational potential energy is the dominant term in the energy budget of the P2 clump. This result is consistent with our mass-to-flux ratio analysis. The kinetic energy is also comparable to the gravitational energy. However, we note that we used our estimated plane-of-sky magnetic field strength to calculate the magnetic energy. Considering 3D magnetic field vectors are randomly oriented, the 3D magnetic field strength can be a factor of 4/$\pi$  larger than the plane-of-sky magnetic field strength \citep{Crutcher2004}.If we applied this correction to our value, the magnetic energy would become 9.8 $\times$ 10$^{46}$ erg, which is still smaller than gravitational energy. However, this is a statistical correction that may not be relevant to this region. Future Zeeman measurements of the magnetic field strength along the line of sight would allow us to determine the 3D magnetic field strengths in the region, and thus calculate more accurate magnetic energy values (e.g., \citealt{Hwang2024}).

\subsection{Scenarios changing magnetic field orientations} \label{sec:sce}

We have shown that the magnetic field orientations on core and clump scales are roughly perpendicular to each other in the P2 clump (Figure \ref{fig:polmag2}).  
The JCMT and ALMA observations show the magnetic fields on different physical scales.
The JCMT observations trace scales from $\sim$2 pc to 0.3 pc and the ALMA observations from $\sim$0.3 pc (the maximum recoverable scale of the ALMA observations of 13$''$) down to 0.02 pc. We can see that there is little to no overlap in size scales: the ALMA observations are probing the magnetic field within the beam of the JCMT. Therefore, we need observations on intermediate scales between those mapped by the JCMT and ALMA to find the size scale on which the transition between magnetic field orientations occurs.

\begin{figure*}[htb!]
\epsscale{0.5}
\plotone{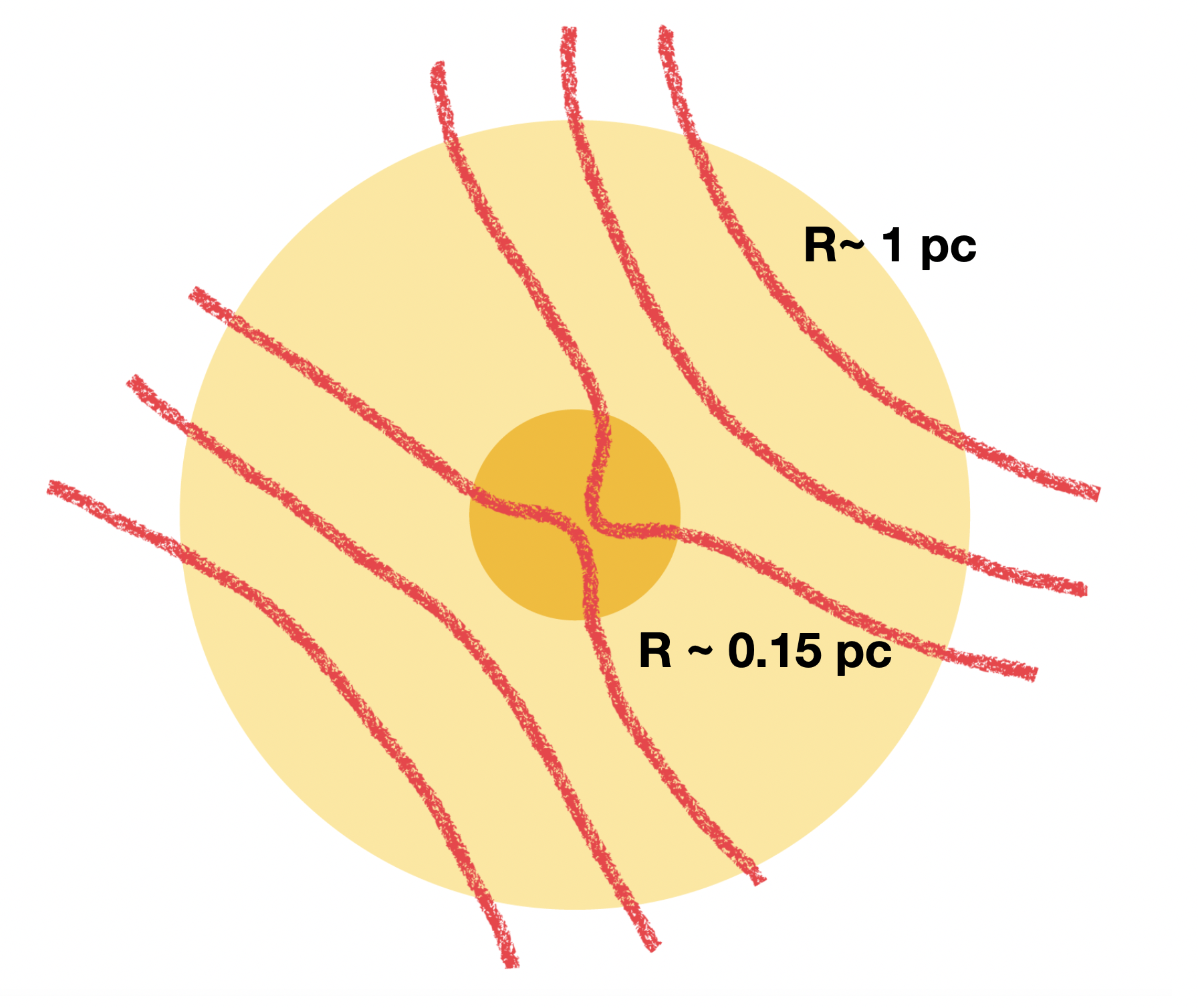}
\caption{A schematic view showing a possible scenario to explain the differing magnetic field orientations on clump (radius $R$ = 1 pc) and core ($R$ = 0.15 pc) scales. Red lines show magnetic field lines. 
\label{fig:view}}
\end{figure*}

We suggest two possible scenarios to explain the change in magnetic field orientations between the clump to the core in G28. In the first scenario, gravitational collapse could change the magnetic field orientations from the P2 clump to the core. In the ambipolar diffusion model, magnetic field lines are aligned along the minor axis of an initially magnetically subcritical clump or core (e.g, \citealt{Mouschovias1999}). In this model, neutral particles in the clump or core decouple from the ions and fall in under gravity.  Once the core becomes magnetically supercritical, the ions drag the magnetic field inwards, creating a characteristic "hourglass" geometry. In a theoretical hourglass model of a collapsing core, the mass-to-flux ratio changes from transcritical to supercritical as the density increases from the edge to the center (e.g., \citealt{Bino2021}). While field lines are only slightly curved on the clump scale, the lines are highly curved on the core scale, where the densities are highest. Figure \ref{fig:mtob} shows that the mass-to-flux ratios we measure are similar to this prediction, transitioning from magnetically subcritical to supercritical from the outer diffuse to the inner dense regions of the P2 clump. We note that the fractional uncertainty on the mass-to-flux ratio is about 33\%, and so the clump could be trans-, rather than super-critical, although as discussed about, there is reason to expect that the clump's supercriticality may in fact be underestimated by our analysis. We would need Zeeman measurements to determine the 3D magnetic field and estimate a more accurate mass-to-flux ratio. The dense core is embedded in the densest part of the P2 clump, in which the magnetic field strength is the highest, but is magnetically supercritical and undergoing gravitational collapse \citep{Liu2020}.
The magnetic field lines are dragged toward the central core, and the field orientations are shifted by 90 degrees with respect to the large-scale field (Figure \ref{fig:view}). \citet{Liu2020} found that the magnetic field revealed by ALMA is preferentially aligned with the direction of gravity in the P2 clump, which further supports this scenario. 

The second scenario posits that the observed magnetic field geometry is the effect of the outflow from the core. \citet{Liu2020} detected an outflow from the core embedded in the P2 clump using ALMA CO (2-1) line observations (see Figure \ref{fig:co} in Appendix \ref{sec:co}). The CO outflow is aligned along the northwest/southeast directions, with a wide opening angle. The IRS 2 source is located in the core observed by the ALMA (R.A. = 18$^h$ 42$^m$ 51.92$^s$, Dec. = -03$^\circ$ 59$'$ 54$''$). However, it cannot be confirmed that the outflow arises from the IRS 2 source due to the differences in their positions. Although the kinetic energy is significant in this region and an outflow is detected, it is hard to constrain the effect of the outflow on the magnetic field orientations. To do so, observing other outflow or dense core tracers would be necessary.
 We cannot rule out the second scenario, but gravitational collapse could be the primary cause of the observed magnetic field geometry based on the mass-to-flux ratios in G28 (Section \ref{sec:mtob}).

\section{Summary}\label{sec:Sum}

We present observations of polarized 850 $\micron$ dust emission performed using the JCMT as part of the JCMT BISTRO Survey, and N2H+ (1-0) line observations performed using the TRAO, of the massive star-forming region G28. We investigate the magnetic field orientations and strengths on clump scales in the P2 clump of G28 using the JCMT observations. We compare the magnetic field orientations with those seen on core scales using archival ALMA data. The magnetic field segments obtained using the JCMT and ALMA are oriented nearly perpendicular to one another in the central parts of the P2 clump. We obtain the distribution of magnetic field strengths using volume density, velocity dispersion, and polarization angle dispersion. 
Using the DCF method, we use these quantities to derive estimated magnetic field strengths which vary from 54 to 432 $\mu$G with a mean value of 185 $\pm$ 65 $\mu$G. The magnetic field is strongest in the center of G28, where a core detected by ALMA is embedded. We also estimated the mass-to-flux ratios to investigate the relative importance of magnetic fields and gravity. We find mass-to-flux ratios ranging from 0.3 to 3.2 with a mean value of 1.5 $\pm$ 0.5. The mass-to-flux ratio is larger than 1 in the central parts of the P2 clump, and decreases with increasing distance from the peak intensity position of the P2 clump. Overall, the clump is magnetically supercritical, which indicates the magnetic field is not large enough to resist gravitational collapse. We also estimate the kinetic, gravitational potential, and magnetic energies of 12 $\times$ 10$^{46}$, 21 $\times$ 10$^{46}$, and 6 $\times$ 10$^{46}$ erg in the P2 clump. The gravitational potential energy is the dominant energy term in the P2 clump. We interpret the perpendicular magnetic field orientations on clump and core scales as potentially resulting from the gravitational collapse of the P2 clump. 

\software{CLASS \citep{Pety2005, gildasteam2013}, KAPPA \citep{Currie2008}, Starlink \citep{Jenness2013}}
\facilities{JCMT}

\begin{acknowledgments}
The JCMT is operated by the East Asian Observatory on behalf of The National Astronomical Observatory of Japan; Academia Sinica Institute of Astronomy and Astrophysics; the Korea Astronomy and Space Science Institute; the Operation, Maintenance and Upgrading Fund for Astronomical Telescopes and Facility Instruments, budgeted from the Ministry of Finance of China. Additional funding support is provided by the Science and Technology Facilities Council of the United Kingdom and participating universities and organizations in the United Kingdom, Canada and Ireland. Additional funds for the construction of SCUBA-2 were provided by the Canada Foundation for Innovation. 
The authors wish to recognize and acknowledge the very significant cultural role and reverence that the summit of Maunakea has always had within the indigenous Hawaiian community.  We are most fortunate to have the opportunity to conduct observations from this mountain. This paper makes use of the following ALMA data: ADS/JAO.ALMA\#2016.1.00248.S and ADS/JAO.ALMA\#2017.1.00793.S. ALMA is a partnership of the ESO (representing its member states), NSF (USA) and NINS (Japan), together with NRC (Canada), MOST and ASIAA (Taiwan), and KASI (Republic of Korea), in cooperation with the Republic of Chile. The Joint ALMA Observatory is operated by ESO, AUI/NRAO, and NAOJ.
K.P. is a Royal Society University Research Fellow, supported by grant number URF\textbackslash{}R1\textbackslash{}211322.
C.W.L. is supported by the Basic Science Research Program through the NRF funded by the Ministry of Education, Science and Technology (NRF- 2019R1A2C1010851) and by the Korea Astronomy and Space Science Institute grant funded by the Korea government (MSIT; project No. 2024-1-841-00).
K. Q. acknowledges National Natural Science Foundation of China (NSFC) grant Nos. 12425304 and U1731237, and the National Key R\&D Program of China with Nos. 2023YFA1608204 and 2022YFA1603103. 
F.P. acknowledges support from the Spanish Ministerio de Ciencia, Innovación y Universidades (MICINN) under grant numbers PID2022-141915NB-C21.
D.J.\ is supported by NRC Canada and by an NSERC Discovery Grant.
W.K. is supported by the National Research Foundation of Korea (NRF) grant funded by the Korea government (MSIT) (RS-2024-00342488). 
M.Tahani is supported by the Banting Fellowship (Natural Sciences and Engineering Research Council Canada) hosted at Stanford University and the Kavli Institute for Particle Astrophysics and Cosmology (KIPAC) Fellowship.
J.K. is supported by the Royal Society under grant number RF\textbackslash{}ERE\textbackslash{}231132, as part of project URF\textbackslash{}R1\textbackslash{}211322.
N.B.N. was funded by Vingroup Innovation Foundation (VINIF) under project code VINIF.2023.DA.057.T
M.Tamura. is supported by JSPS KAKENHI grant No.24H00242. J.K. is supported by JSPS KAKENHI grant No.24K07086.
E.C. acknowledges the financial support from grant RJF/2020/000071 as a part of the Ramanujan Fellowship awarded by the Science and Engineering Research Board (SERB), Department of Science and Technology (DST), Govt. of India.
T.L. acknowledges the supports by the National Key R\&D Program of China (No. 2022YFA1603100), National Natural Science Foundation of China (NSFC) through grants No.12073061 and No.12122307, and the Tianchi Talent Program of Xinjiang Uygur Autonomous Region.
S.-P.L. acknowledges the grants from the National Science and Technology Council of Taiwan under project numbers 112-2112-M-007-011 and 113-2112-M-007-004.
R.S.F. was supported by the Visiting Scholars Program provided by the NAOJ Research Coordination Committee, NINS (NAOJ-RCC-23DS-050).
\end{acknowledgments}

\appendix

\section{Data reduction procedures}\label{sec:datareduce}

The raw data were reduced using the Submillimetre User Reduction Facility (SMURF) package \citep{Chapin2013} from the Starlink software suite \citep{Currie2014}. The $pol2map${\footnote{\url{http://starlink.eao.hawaii.edu/docs/sun258.htx/sun258ss75.html}}} command is run twice to reduce the POL-2 data. In the first run of $pol2map$, raw bolometer timestreams for each observation are converted into separate Stokes $I$, $Q$, and $U$ timestreams. In this process, the SMURF routine $makemap$ \citep{Chapin2013} produces individual Stokes $I$ maps from the Stokes $I$ timestreams and then coadds those maps to form an initial Stokes $I$ map. The initial Stokes $I$ maps are used to make masks defining areas of astrophysical emission. The second run of $pol2map$ creates final Stokes $I$, $Q$, and $U$ maps, and a polarization segment catalog. In this second run, we used the $skyloop$ and $mapvar$ parameters in $pol2map$. The $skyloop$ parameter improves the signal-to-noise ratio and image fidelity by reducing all 45 observations concurrently. The $mapvar$ parameter calculates the variances of the Stokes $I$, $Q$, and $U$ maps from the spread of pixel data values between the individual observations. We used the August 2019 instrumental polarization model to correct the Stokes $Q$ and $U$ maps. 

\section{N$_2$H$^+$} \label{sec:n2h+}

\begin{figure*}[htb!]
\epsscale{1.0}
\plotone{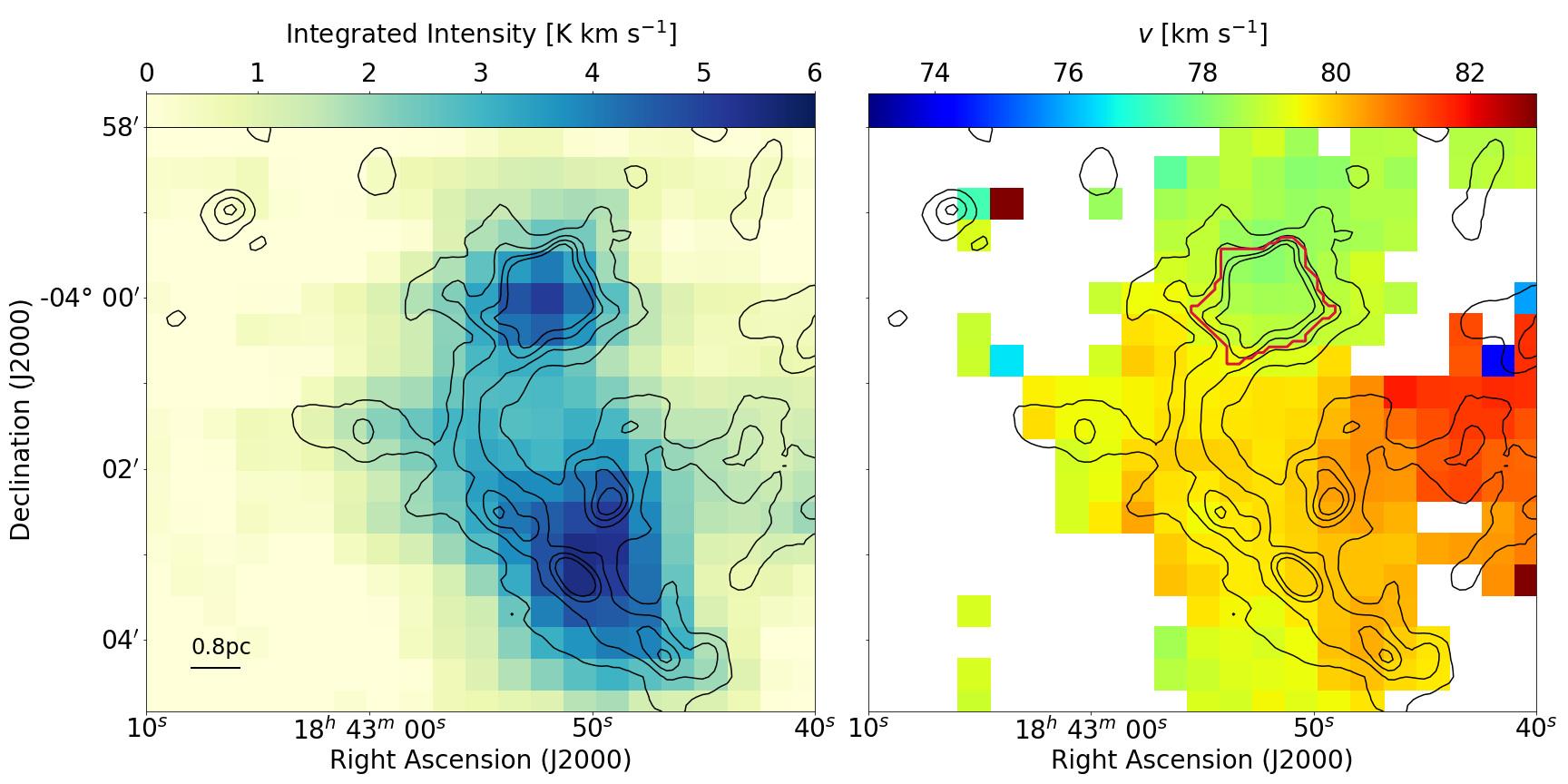}
\caption{Maps of integrated intensity (left) and the centroid velocity (right) of N$_2$H$^+$ obtained by the TRAO observations. Contours in both panels are the same in Figure \ref{fig:col}. 
\label{fig:moments}}
\end{figure*}

Figure \ref{fig:moments} shows maps of integrated intensity and centroid velocity of N$_2$H$^+$ obtained by the TRAO observations. We integrated N$_2$H$^+$ data from 60 to 110 km s$^{-1}$. The centroid velocity is determined from the Gaussian fit value of the isolated component (Section \ref{sec:veldisp}).

\section{CO outflow} \label{sec:co}

\begin{figure*}[htb!]
\epsscale{1.0}
\plotone{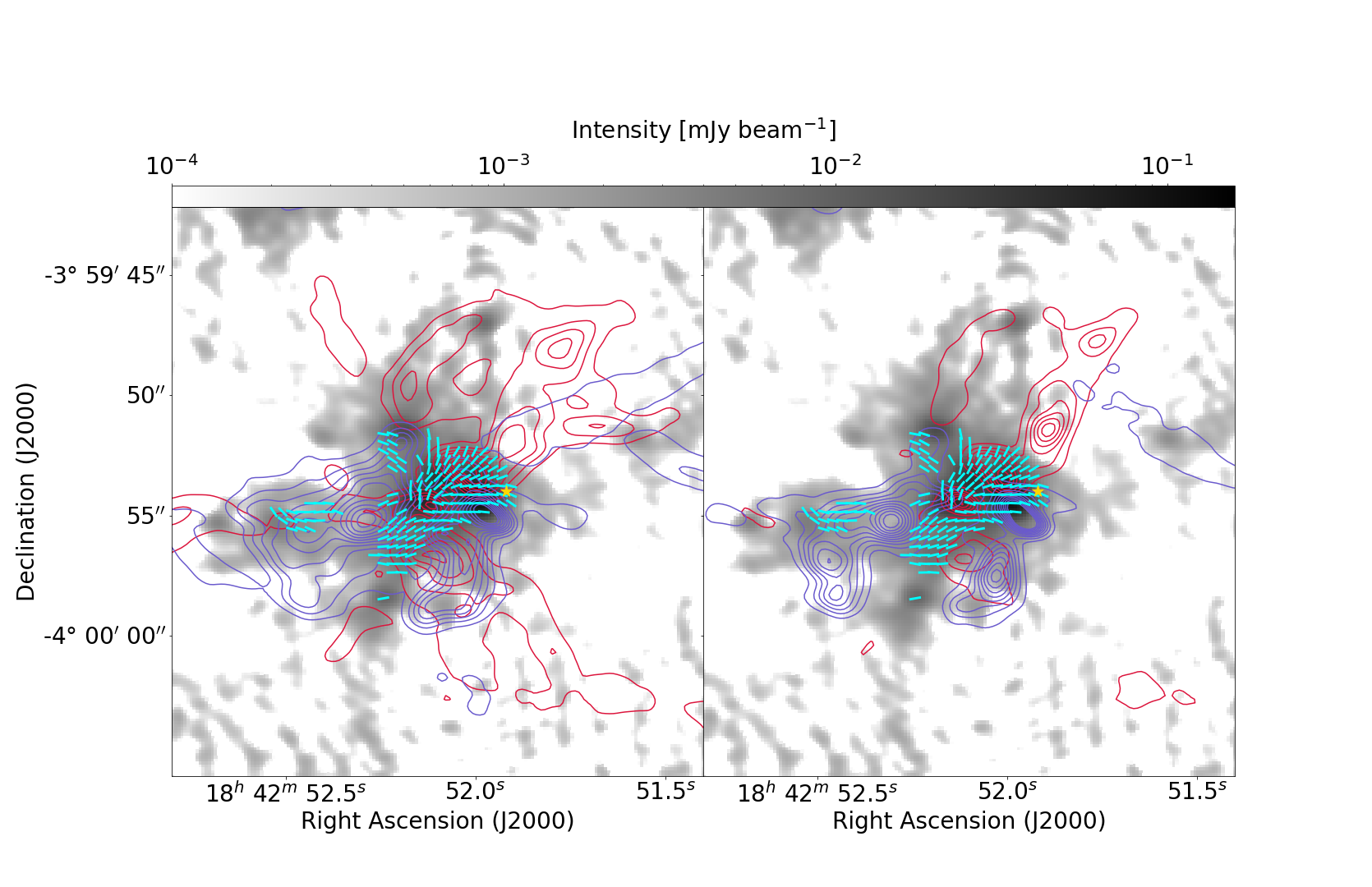}
\caption{Maps of molecular outflows detected in CO (2-1) by the ALMA observations. The cyan segments show magnetic field orientations obtained by the ALMA. In the left panel, CO emission is integrated from 45 to 74 km s$^{-1}$ for the blue robe, and from 84 to 115 km s$^{-1}$ for the red robe. In the right panel, high-velocity CO emissions are integrated from 45 to 59 km s$^{-1}$ for the blue robe, and from 99 to 115 km s$^{-1}$ for the red lobe. The blue and red contour levels indicate 5, 15, 25, 35, 45, 55, 65, 75, 85, and 95 \% of the integrated values. The yellow stars in both panels indicate the location of IRS 2 source.
\label{fig:co}}
\end{figure*}

Figure \ref{fig:co} shows the blueshifted and redshifted CO outflows obtained by the ALMA observations \citep{Liu2020}. Magnetic field orientations are overlapped in the figure. The yellow star in the figure is the location of IRS 2 source. It is hard to constrain outflow direction from the IRS 2 source.




\end{document}